 \documentclass[final,3p,times,authoryear]{elsarticle}
\usepackage{amssymb}
\usepackage{lipsum}
\usepackage{booktabs}
\usepackage[table]{xcolor}
\usepackage{threeparttable}
\usepackage{rotating}
\usepackage{tabularray}

\definecolor{headerblue}{RGB}{79,129,189}
\definecolor{sectionblue}{RGB}{184,204,228}
\definecolor{lightrow}{RGB}{242,242,242}

\usepackage{pdflscape}

\usepackage{longtable}

\usepackage{fancyhdr}
\usepackage[colorlinks=true]{hyperref}

\usepackage{caption}
    \labelformat{figure}{%
        {%
            \color{blue}%
            Fig.~#1%
        }%
    }
    \labelformat{table}{%
        {%
            \color{blue}%
            Table~#1%
        }%
    }

\usepackage{lineno}

\defcitealias{unitednationsofficeforouterspaceaffairsunoosa2021}
{United Nations Office for Outer Space Affairs (UNOOSA), 2021}

\defcitealias{europeanspaceagency2025}
{European Space Agency (ESA), 2025}

\journal{Journal of Cleaner Production}

\begin{document}

\begin{frontmatter}

%% Title, authors and addresses

\title{Conceptualizing and Defining the Circular Space Economy}

\author[1,2]{Jonas Bahlmann}\corref{cor1}
\cortext[cor1]{Corresponding author.}
\ead{jonas.bahlmann@uni.lu}

\author[3]{Michael Saidani}
\ead{michael.saidani@etsmtl.ca}

\author[2]{Enrico Stoll}
\ead{e.stoll@tu-berlin.de}

\author[1]{Andreas M. Hein}
\ead{andreas.hein@uni.lu}

\affiliation[1]{organization={Space Systems Research Group, Interdisciplinary Centre of Security, Reliability and Trust (SnT), University of Luxembourg},
            addressline={29 Av. J. F. Kennedy}, 
            postcode={1855},
            city={Luxembourg},
            country={Luxembourg}}

\affiliation[2]{organization={Chair of Space Technology, Technical University Berlin},
            addressline={Marchstrasse 12-14}, 
            postcode={10587},
            city={Berlin},
            country={Germany}}
            
\affiliation[3]{organization={Department of Systems Engineering, École de Technologie Supérieure, University of Quebec},
            addressline={1100 Notre-Dame Street West}, 
            postcode={H3C 1K3},
            city={Montréal},
            country={Canada}}

\begin{abstract}

Space faces significant sustainability issues including orbital congestion and debris accumulation. The continued growth of space operations, accelerated by advancements such as reusable launch systems, further intensifies these pressures. Current mitigation strategies, such as deorbiting spacecraft or transferring them to graveyard orbits, remain inherently linear. This “take–make–waste” approach is environmentally unsustainable and economically inefficient. On Earth, similar challenges have driven the development of the circular economy (CE), which aims to eliminate waste and pollution, circulate resources at their highest value, and decouple economic growth from finite resource consumption. While these objectives have been extensively studied across terrestrial sectors, their application to the space domain remains largely unexplored. In particular, the concept of a circular space economy (CSE) remains constrained by narratives centered on reuse, recycling, and in-orbit servicing, lacking a structured definition, consistent terminology, and a clearly defined, comprehensive scope. This lack complicates the systematic integration of circularity into mission design, policy frameworks, and space system architectures. After a detailed analysis of established CE definitions and CSE definition proposals, this work conceptualizes the CSE and introduces a structured definition for the first time. It analyzes Earth-space distinctions, clarifies the relationship between space sustainability and the CSE, establishes the 10R Space Framework to narrow, slow, and close resource loops, and distinguishes three operational environments: (I) the CE in space, (II) the CE of the terrestrial (space) sector, and (III) the CE of celestial bodies beyond Earth. Ultimately, this work enables a shared understanding among stakeholders and aims to strengthen the concept’s recognition in the space sustainability debate.

\end{abstract}

%%Graphical abstract
%\begin{graphicalabstract}
%\includegraphics{grabs}
%\end{graphicalabstract}

%%Research highlights
%\begin{highlights}
%\item Research highlight 1
%\item Research highlight 2
%\end{highlights}

\begin{keyword}
%% keywords here, in the form: keyword \sep keyword, up to a maximum of 6 keywords
Circular space economy \sep Circular economy \sep Definition \sep Space sustainability \sep Sustainable development \sep Groundwork

%% PACS codes here, in the form: \PACS code \sep code

%% MSC codes here, in the form: \MSC code \sep code
%% or \MSC[2008] code \sep code (2000 is the default)

\end{keyword}

\end{frontmatter}

%\tableofcontents

% \linenumbers

%% main text

%\fancyhf{} -> Deletes previous Style
\setlength{\footskip}{20pt} %-> Adds space between text and bar
\fancyhead[L]{\textit{Preprint | CC BY-NC-ND 4.0}}
\fancyhead[R]{\textit{May 14, 2026}}

\renewcommand{\headrulewidth}{0.4pt}
\renewcommand{\footrulewidth}{0.4pt}

\section{Introduction}
\label{sec:introduction}

“Houston, we have a problem”: more than 70\% of tracked objects in Earth orbit are debris, while an estimated 1.2 million untracked artificial fragments between $1$
 $\mathrm{cm}$ and $10$ $\mathrm{cm}$ pose an additional collision risk, leading to more collision avoidance maneuvers and ultimately higher costs (\citetalias{europeanspaceagency2025}). This situation has emerged because, since the beginning of the space age in 1957, space activities have largely followed a linear “take–make–waste” model, with limited consideration of end-of-life (EoL) strategies. As space activities rapidly expand \citep{europeanspaceagency2025} -- driven by increasing launch rates, new actors, and reduced launch costs through reusable launch systems -- the density of operational spacecraft in heavily populated orbits is approaching that of debris. This increases the likelihood of cascading collision events, commonly referred to as the Kessler Syndrome \citep{kessler1978}.

\begin{table}[t!]
	\setlength\extrarowheight{0pt} % for a bit of visual "breathing space"
	\centering
    \footnotesize
 	%\caption{List of acronyms and abbreviations}
	\label{tab:list_of_acronyms_and_abbreviations}
    \begin{tabular}{|p{0.17\textwidth}p{0.42\textwidth}|}
		\hline
        \multicolumn{2}{|c|}{}                                                                                                                   \\
        \multicolumn{2}{|l|}{\textbf{List of acronyms and abbreviations}}                                                                        \\
        \multicolumn{2}{|c|}{}                                                                                                                   \\
        %AOCS                                  & Attitude and Orbit Control System                                                                \\
        CE                                    & Circular economy                                                                                 \\
        CEiS                                  & Circular economy in space                                                                        \\
        %CEO                                   & Chief Executive Officer                                                                          \\
        %CONFERS                               & Consortium for Execution of Rendezvous and Servicing Operations                                  \\
        %COPUOS                                & Committee on the Peaceful Uses of Outer Space                                                    \\
        CSE                                   & Circular space economy                                                                           \\
        %DLR                                   & Deutsches Zentrum für Luft- und Raumfahrt (en: German Aerospace Center)                          \\
        EoL                                   & End-of-life                                                                                      \\
        ESA                                   & European Space Agency                                                                            \\
        GDP                                   & Gross domestic product                                                                           \\
        GEO                                   & Geosynchronous Orbit                                                                             \\
        %GNC                                   & Guidance, Navigation, and Control                                                                \\
        ISO                                   & International Organization for Standardization                                                   \\
        %ISRU                                  & In-situ Resource Utilization                                                                     \\
        %ISS                                   & International Space Station                                                                      \\
        %ITU                                   & International Telecommunication Union                                                            \\
        %L                                     & Sun-Earth Lagrange Point                                                                                   \\
        %LCA                                   & Life Cycle Assessment                                                                            \\
        LEO                                   & Low Earth orbit                                                                                  \\
        %MEV                                   & Mission Extension Vehicle                                                                        \\
        %MS                                    & Microsoft                                                                                        \\
        %NASA                                  & National Aeronautics and Space Administration                                                    \\
        %OSAM                                  & On-Orbit Servicing, Assembly, and Manufacturing                                                  \\        
        %OSIRIS-APEX                           & Origins, Spectral Interpretation, Resource Identification, and Security – Apophis Explorer       \\
        %OSIRIS-REx                            & Origins, Spectral Interpretation, Resource Identification, and Security – Regolith Explorer      \\
        R-frameworks (e.g., 3R)      & Structured sets of circular actions (e.g., reduce, reuse, recycle) \\
        %SBSP                                  & Space-based Solar Power                                                                          \\
        %SPS                                   & Solar Power Satellites                                                                           \\
        %TRL                                   & Technology Readiness Level                                                                       \\
        %UK                                    & United Kingdom                                                                                   \\
        %UN                                    & United Nations                                                                                   \\
        UNOOSA                                & United Nations Office for Outer Space Affairs                                                    \\
        %US                                    & United States of America                                                                         \\
                                              &                                                                                                  \\        
        \hline
    \end{tabular}
\end{table}

These developments threaten the long-term usability of the space environment and call into question the environmental, social, and economic sustainability of current space activities. In this context, the concept of space sustainability provides an overarching framework to ensure the safe and beneficial use of outer space while preserving access for current and future generations (\citetalias{unitednationsofficeforouterspaceaffairsunoosa2021}; \citealp{wilson2023}). Although current mitigation strategies such as space situational awareness, collision avoidance maneuvers, spacecraft design adaptations, passive and active debris removal, graveyard orbits, and deorbiting measures represent important capabilities to develop and maintain, they primarily aim to prolong the unsustainable status quo rather than enable a radical systemic shift in how space missions are conducted.

Comparable challenges on Earth have led to the development of the circular economy (CE), a concept aimed at eliminating waste and pollution, circulating resources at their highest value, and decoupling economic growth from the consumption of finite resources \citep{kirchherr2023, ellenmcarthurfoundation2026}. The CE provides a framework for slowing, narrowing, and closing resource loops \citep{bocken2016a} and has been applied across a wide range of sectors. However, only 6.9\% of the global economy is estimated to be circular \citep{circleeconomy2025}, leaving an enormous “€25.4 trillion (± €4.7 trillion) in avoidable annual economic value lost to linear material use, equivalent to almost 31\% of global GDP [gross domestic product]” \citep{circleeconomy2026}.

The implementation of CE strategies is often structured through so-called “R-frameworks,” which categorize circular actions at different levels. Among these, the 9R framework is widely recognized as a comprehensive classification \citep{kirchherr2017}. Other frameworks, such as the 3R \citep{manickam2019}, the 10R \citep{reike2018}, and the Ellen MacArthur Foundation’s “Butterfly Diagram” \citep{ellenmcarthurfoundationn.d.}, vary in scope and complexity. Nevertheless, the concept of CE is still developing. A recent study documented 221 distinct definitions of the CE \citep{kirchherr2023}, highlighting its conceptual width. With its origins in the 1960s and 1970s \citep{winans2017}, the CE became the subject of an International Organization for Standardization (ISO) standard only decades later \citep{iso2024c}. Despite the abundance of CE definitions, no universally accepted definition has emerged, reflecting the difficulty practitioners face in establishing a conceptual basis. It is therefore reasonable to assume a similar evolutionary process in defining the circular space economy (CSE), which this paper seeks to initiate.

While some space missions have demonstrated the feasibility of circular actions to a limited extent, broader benefits are expected once the CSE has fully unfolded, including increased resource efficiency, extended system lifetimes, reduced debris generation, lower long-term costs, and reduced environmental impacts both in space and on Earth \citep{bahlmann2024}. In this context, circularity -- understood as the degree of alignment with CE principles such as systems thinking, value creation, value sharing, resource stewardship, resource traceability, and ecosystem resilience \citep{iso2024c} -- is increasingly considered a necessity for achieving long-term space sustainability and ensuring future operability by institutions such as ESA \citep{europeanspaceagency2023}.

Despite its current strong momentum, the concept of a CSE remains insufficiently defined. Existing contributions rely on fragmented or context-specific definitions, which are referred to in this work as CSE definition proposals to distinguish them analytically from established CE definitions. These proposals are often embedded within broader studies and show limited convergence in terminology and scope \citep{leonard2023, wilson2023, saadetturnbull2026}. Current narratives are predominantly centered on reuse, recycling, and in-orbit servicing \citep{europeanspaceagency2023, jah2024a, bennett2025}, thereby capturing only a subset of the range of possible circular actions \citep{iso2024c}. Moreover, a systematic distinction between in-space and terrestrial space-related activities is absent. This lack of a shared understanding complicates the integration of circularity into mission design, policy frameworks, and space system architectures.

Hence, the objective of this work is to conceptualize and subsequently define the CSE in a structured approach to enable a shared understanding across stakeholders. In particular, it addresses the following research questions: (I) How can the CSE be defined, and what key elements constitute such a definition? (II) How can the CSE be conceptualized, and which elements and relationships shape a corresponding framework?

To address these questions, the paper is structured as follows. Section \ref{sec:methodology} presents the selection principles for the analyzed definitions, the development of the coding framework, and the overall methodology. Section \ref{sec:analysis} investigates conceptual patterns of established CE definitions and CSE definition proposals based on the derived coding framework. Section \ref{sec:cse_conceptualization} conceptualizes the CSE to address identified gaps. Synthesizing all insights from the previous sections, Section \ref{sec:cse_definition} presents a structured CSE definition for the first time. Subsequently, Section \ref{sec:discussion_and_conclusion} discusses the results, provides final remarks on this work, along with directions for future research.

\section{Materials and methods}
\label{sec:methodology}

\begin{figure}[ht]
	\centering 
	\includegraphics[width=0.9\textwidth]{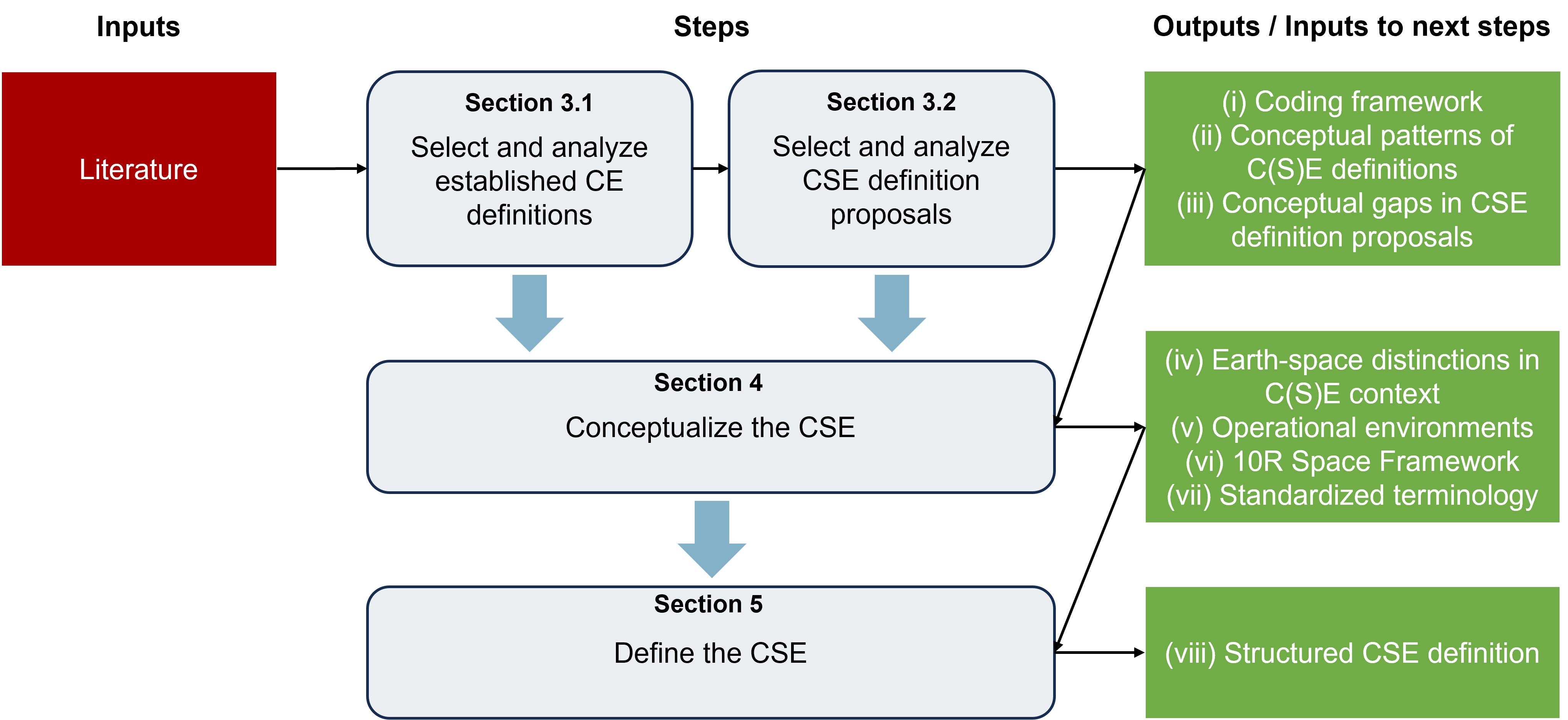}	
	\caption{Methodological flowchart.} 
	\label{fig:methodological_flowchart}
\end{figure}

This work conceptualizes the CSE and subsequently develops a structured definition through the methodology shown in \ref{fig:methodological_flowchart}. It combines a staged literature-based analysis with original conceptual development. First, established CE definitions were selected and analyzed to identify recurring conceptual patterns and coding dimensions. Second, existing CSE definition proposals were selected and analyzed to validate the coding framework and identify recurring as well as missing conceptual elements. These two analytical steps revealed insights and gaps that were addressed in the conceptualization of the CSE. Finally, all results were consolidated into a structured CSE definition.

\subsection{Literature selection}
\label{sec:methodology_literature_selection}

The sample development was conducted in two streams: (I) established CE definitions and (II) CSE definition proposals. Both peer-reviewed articles and grey literature such as official reports and institutional websites were considered, given that the CE domain has also been substantially developed outside academia \citep{kirchherr2017}. This applies even more to the CSE domain. In three cases, no publication date could be identified; these sources were nevertheless retained because they provide analytically relevant formulations, either through institutional relevance in the CE sample or by contributing to exhaustive coverage in the CSE sample.

(I) In this context, established CE definitions (also referred to as CE definitions for readability) refer to such that have gained relevance through frequent citation, institutional adoption, standardization processes, or sustained use in academic, policy, or practice-oriented discourse. Further objectives were to select CE definitions that capture the evolution of the CE concept and reflect a balance between academic and institutional considerations, for instance.

The initial CE sample was developed from CE definition reviews, particularly \citet{kirchherr2017} and \citet{kirchherr2023}, and then complemented through targeted searches in Google Scholar, Scopus, and Google Search using the keywords \{circular economy OR CE\} AND \{definition, concept, principle, framework, standard, policy\}. Saturation was reached after ten CE definitions: additional contributions did not alter the coding framework or strengthen the analytical basis of the study. The selection covers definitions from 2013 onward to balance sources that have had sufficient time to become established (e.g., through citation count), with more recent definitions that have already gained relevance through institutional use.

(II) The selection of CSE definition proposals required a broader search strategy. Unlike CE, the CSE is not yet an established research field with a consolidated terminology or a dedicated body of peer-reviewed literature on its definition. Consequently, a conventional database search including terms such as “definition” would have been insufficient, as relevant contributions frequently describe, frame, or imply the CSE without explicitly defining it.

Therefore, the sample was developed through a targeted search in Google Scholar, Scopus, and Google Search using the keywords \{circular space economy OR CSE OR space circular economy OR circular economy in space OR CE in space OR circular economy for space OR CE for space OR circularity in space OR circularity for space OR space circularity\}. The search was complemented by snowballing \citep{wohlin2014} and expert-informed source identification based on previous studies \citep{bahlmann2024}. The resulting publications were skimmed to identify passages containing CSE definition proposals, where present, and the analysis was limited to these core formulations to ensure comparability across heterogeneous sources and avoid coding broader contextual information beyond the scope of this study.

The search identified 13 CSE definition proposals published between 2023 and 2026, reflecting the recent development of the body of literature. We exceeded the ten-definition count used for the CE sample, as each additional CSE definition proposal contributed distinct analytical value given the overall diversity in content and scope. Although the selection aims to be comprehensive, completeness cannot be guaranteed due to the dispersed and inconsistent use of CSE-related terminology.

\subsection{Coding framework}
\label{sec:methodology_coding_framework}

To develop a CSE definition, it is necessary to understand how CE definitions are structured. The selected CE definitions were sorted by year of publication, where available, and analyzed for recurring conceptual patterns. Drawing inspiration from \citet{kirchherr2023}, a coding framework was developed through an iterative process around five dimensions: core concept, aim, enabler, beneficiary, and framework / conceptual model. Each dimension was triggered by one respective guiding question: What is CE primarily described as? What does it aim to achieve? Which mechanisms enable it? Who benefits from it? Is an explicit framework or conceptual model provided?

The framework was designed to remain as accessible as possible while being sufficiently detailed to classify all elements of each definition across terrestrial and space contexts. Coding remained close to the wording of the definitions, avoided interpretation and multiple classifications where possible, and considered surrounding text only when necessary to clarify meaning. The coding framework was subsequently applied to CSE definition proposals to validate its suitability beyond established CE definitions, requiring no structural deviations while enabling the identification of recurring and missing elements in the emerging discourse.

\section{Analysis of the current definition landscape}
\label{sec:analysis}

The objective of this section is to examine how established CE definitions are composed and to assess whether CSE definition proposals show comparable conceptual depth. The definition selection principles, coding framework development, and coding process are specified in Section \ref{sec:methodology}.

\subsection{Established circular economy definitions}
\label{subsec:analysis_ce_def}

Despite considerable variation in wording, CE definitions are relatively consistent in both their structure and conceptual orientation. As shown in \ref{tab:CE_defintions_coded}, all selected definitions include a core concept and an aim, while almost all include enablers and a framework or conceptual model.

\begin{table}[ht]
\centering
\footnotesize
\caption{Coding framework applied to ten CE definitions, sorted by year of publication (if available) and selected according to the principles in Section \ref{sec:methodology}. Coding dimensions are partly adapted from \citet{kirchherr2023}. Detailed coding results are provided in \ref{appendix_a_detailed_analysis_ce_definitions}.}
\label{tab:CE_defintions_coded}
\begin{tabular}{>{\raggedright\arraybackslash}p{5.2cm} c c c c >{\centering\arraybackslash}p{2.4cm}}
\toprule
\textbf{Source} 
& \textbf{Core concept} 
& \textbf{Aim} 
& \textbf{Enabler} 
& \textbf{Beneficiary} 
& \textbf{Framework / Conceptual model} \\
\midrule
\cite{ellenmcarthurfoundation2013} & X & X & X  & -- & X  \\
\cite{geissdoerfer2017}            & X & X & X  & -- & X  \\
\cite{murray2017}                  & X & X & X  & X  & X  \\
\cite{kirchherr2017}               & X & X & X  & X  & X  \\
\cite{unenvironmentprogramme2019}  & X & X & X  & -- & X  \\
\cite{kirchherr2023}               & X & X & X  & X  & X  \\
\cite{iso2024c}                    & X & X & X  & -- & -- \\
\cite{mckinseyandcompany2024}      & X & X & X  & -- & X  \\
\cite{ellenmcarthurfoundation2025} & X & X & X  & X  & X  \\
\cite{europeanunionn.d.}           & X & X & -- & -- & X  \\
\bottomrule
\end{tabular}
\end{table}

The core concept is most often framed as a system or economic system, for example by \citet{kirchherr2017}, \citet{kirchherr2023}, \citet{iso2024c}, and \citet{europeanunionn.d.}, while other definitions highlight restorative, regenerative, or resilient system characteristics \citep{ellenmcarthurfoundation2013, geissdoerfer2017, ellenmcarthurfoundation2025}. Across the selected definitions, aims repeatedly include resource efficiency, value retention, waste reduction, sustainable development, and the decoupling of economic activity from finite resource consumption \citep{unenvironmentprogramme2019, iso2024c, mckinseyandcompany2024}. Enablers are frequently linked to design, business models, technological innovation, and the management of resource flows \citep{geissdoerfer2017, murray2017, kirchherr2023}. 

Frameworks and conceptual models are also commonly present. These include, for instance, the distinction between technical and biological cycles in \citet{ellenmcarthurfoundation2013} and 4R frameworks in \citet{kirchherr2017}, \citet{unenvironmentprogramme2019}, \citet{kirchherr2023}, \citet{ellenmcarthurfoundation2025}, and \citet{europeanunionn.d.}. Beneficiaries are less consistently addressed, with no clear pattern observed in whether they are included. Where specified, examples include humans, current and future generations, and broader benefits for business, people, and the environment \citep{murray2017, kirchherr2017, ellenmcarthurfoundation2025}.

The detailed analysis in \ref{appendix_a_detailed_analysis_ce_definitions} shows that all elements of the selected definitions could be classified, which validates the coding framework for the purpose of this study. Interestingly, the literature review revealed that, even several decades after the emergence of the CE concept, no universally accepted definition exists; instead, conceptual coherence emerges through recurring elements across different sources.

\subsection{Circular space economy definition proposals}
\label{subsec:analysis_cse_def_proposals}

The analysis shows a less mature and less coherent definition landscape. As shown in \ref{tab:CSE_defintions_coded}, all selected proposals include a core concept and an aim, while enablers are only partly addressed. Beneficiaries are rarely specified, and explicit frameworks or conceptual models are almost entirely absent. The detailed classification in \ref{appendix_b_detailed_analysis_cse_definition_proposals} reveals that all textual elements could be classified using the previously developed coding framework.

The core concepts of the CSE definition proposals are predominantly centered on reuse, recycling, and in-orbit servicing \citep{europeanspaceagency2023, jah2024a, bennett2025}, while the broad range of circular actions to close resource loops \citep{iso2024c} is underrepresented. The aims provide a wider perspective and range from minimizing waste, reducing debris generation, and optimizing resource use \citep{wilson2023, jah2024a, saadetturnbull2026} to lowering environmental impacts and supporting long-term space sustainability \citep{europeanspaceagency2023, noaaofficeofspacecommerce2025, saadetturnbull2026}.

Enablers are addressed unevenly across the proposals. Frequently mentioned examples include design, in-orbit servicing and manufacturing, modularity, policy, technological innovation, and business models \citep{wilson2023, bennett2025, d-orbit2026}. Beneficiaries are much less visible and are explicitly mentioned only in three cases, referring to humans in \citet{d-orbit2026} and future generations in \citet{turner2024} and \citet{saadetturnbull2026}. Frameworks or conceptual models are largely absent; only \citet{wilson2023} outline framework-like requirements, without specifying an actual framework or conceptual model.

It is noteworthy that \citet{saadetturnbull2026} specify a 10R framework adapted from \citet{kirchherr2017} and \citet{potting2017} in their report, but do not include it in their working definition. Similarly, contributions such as \citet{yang2025}, \citet{bonifazi2026}, and \citet{schultz2026} provide relevant conceptual elements but no original CSE definitions; therefore, they are excluded from the analysis given the scope of this study.

\begin{table}[ht]
\centering
\footnotesize
\caption{Coding framework applied to 13 CSE definition proposals, sorted by year of publication (if available) and selected according to the principles in Section \ref{sec:methodology}. Coding dimensions are partly adapted from \citet{kirchherr2023}. Detailed coding results are provided in \ref{appendix_b_detailed_analysis_cse_definition_proposals}.}
\label{tab:CSE_defintions_coded}
\begin{tabular}{>{\raggedright\arraybackslash}p{5.2cm} c c c c >{\centering\arraybackslash}p{2.4cm}}
\toprule
\textbf{Source} 
& \textbf{Core concept} 
& \textbf{Aim} 
& \textbf{Enabler} 
& \textbf{Beneficiary} 
& \textbf{Framework / Conceptual model} \\
\midrule
\cite{wilson2023}                       & X & X & X  & -- & X/-- \\
\cite{leonard2023}                      & X & X & X  & -- & --   \\
\cite{europeanspaceagency2023}          & X & X & -- & -- & --   \\
\cite{jah2024a}                         & X & X & X  & -- & --   \\
\cite{bahlmann2024a}                    & X & X & X  & -- & --   \\
\cite{turner2024}                       & X & X & X  & X  & --   \\
\cite{dailey2024}                       & X & X & -- & -- & --   \\
\cite{noaaofficeofspacecommerce2025}    & X & X & -- & -- & --   \\
\cite{bennett2025}                      & X & X & X  & -- & --   \\
\cite{d-orbit2026}                      & X & X & X  & X  & --   \\
\cite{saadetturnbull2026}               & X & X &  X & X  & --   \\
\cite{prismsustainabilitydirectoryn.d.} & X & X & -- & -- & --   \\
\cite{wealthformulan.d.}                & X & X & -- & -- & --   \\
\bottomrule
\end{tabular}
\end{table}

Beyond the uneven coverage of coding dimensions and missing elements, the analysis reveals broader gaps. In particular, a systematic distinction between in-space and terrestrial space-related activities is absent, including a life-cycle perspective from design and manufacturing to end-of-use. None of the identified proposals systematically distinguishes between Earth and space as operational environments, although such a distinction is relevant for scope definition, tracking and management of resource flows, and implementation. The peer-reviewed literature base also remains limited, with only three definition proposals identified \citep{wilson2023, leonard2023, bennett2025}. In addition, the terminology varies significantly across sources, as indicated in Section \ref{sec:methodology}. For instance, the terms space circular economy, circular space economy, circular economy in space, and circular economy for space systems are used interchangeably in \citet{saadetturnbull2026}.

The identified gaps and inconsistencies highlight the need for conceptual clarity; accordingly, they will be addressed in the following section.

\section{Conceptualizing the circular space economy}
\label{sec:cse_conceptualization}

\subsection{Identification of Earth-space distinctions in a circular economy context}
\label{subsec:cse_conceptualization_earth_space_distinctions}

To contextualize the CSE, Earth–space distinctions were examined from today’s operational and technological perspective. The objective was to identify how practical aspects relevant to circularity in space differ from the terrestrial context, rather than to suggest that CE methodologies themselves are fundamentally distinct. Despite the obvious environmental differences, this analysis is relevant for operational and accounting reasons, as the life cycle of terrestrial systems takes place on Earth, whereas the life cycle of space systems is split across both environments: design and manufacturing occur on Earth, deployment and operation span Earth and space, end-of-use takes place in space, and EoL occurs across Earth and space.

As shown in \ref{tab:earth_space_comparison}, Earth offers high circularity potential due to large in-use material stocks and established biological and technical loops. Space, by contrast, is currently characterized by negligible circularity, limited resource accessibility, low object volumes, unclear ownership conditions, and a challenging geopolitical context. At the same time, high unit costs, abundant solar energy, and the accumulation of space debris create strong incentives for circular approaches. This makes the CSE particularly relevant for future large space systems, such as solar power satellites, sunshades, space data centers, and large constellations, where growing in-space material stocks may render circularity increasingly necessary.

\begin{table}[t]
\centering
\footnotesize
\begin{threeparttable}
\caption{Comparison of Earth and space as two distinct environments across key dimensions relevant to circularity. The assessment reflects the current state of technological and operational practice.}
\label{tab:earth_space_comparison}

\begin{tabular}{>{\raggedright\arraybackslash}p{4.3cm}
                >{\raggedright\arraybackslash}p{5.45cm}
                >{\raggedright\arraybackslash}p{5.45cm}}
\toprule

{\textbf{Classification}} &
{\textbf{Terrestrial context}} &
{\textbf{Space context}} \\
\midrule

\multicolumn{3}{l}{\textbf{General}} \\
\rowcolor{lightrow}Economy type & Primarily linear (<10\% circular\textsuperscript{1}) & Linear (circularity negligible) \\
Circular loop viability & Biological (renewable) \& technical (finite res.) & Biological (controlled environment) \& technical \\
\rowcolor{lightrow}Circularity potential & High in-use material stocks enable CEs of scale & High unit costs (+), low object volume and in-use material stocks (-) \\
Life cycle location & Full life cycle occurs on Earth & Environmental split; Design \& manufacturing: Earth; deployment \& operation: Earth/space; end-of-use: space; EoL: Earth/space \\

\multicolumn{3}{l}{\textbf{Resources \& Energy}} \\
\rowcolor{lightrow}Resource types & Biological, technical, digital & Primarily technical; digital \\
Resource accessibility & High & Low (celestial bodies \& artificial objects) \\
\rowcolor{lightrow}Energy availability & High & High (solar) \\
Waste & Poses high risk (e.g., pollution) & Poses high risk (e.g., Kessler Syndrome) \\

\multicolumn{3}{l}{\textbf{Environmental \& Technical}} \\
\rowcolor{lightrow}Environmental conditions & Favorable to unfavorable (depending on location) & Thermal challenges, vacuum, microgravity \\
Required technology & Largely developed and tested & Partly developed, largely untested in space \\

\multicolumn{3}{l}{\textbf{Legal \& Political}} \\
\rowcolor{lightrow}Legal context (regulation) & Sector-/industry-specific; governed by national and international laws; partly unregulated domains (e.g., high seas) & International treaties (not binding); unregulated \\
Ownership (territory/resources/waste) & Private, public, common heritage & Province of all mankind\textsuperscript{2}; unclear \\
\rowcolor{lightrow}Geopolitical context & Sovereign territories & Shared global common \\
\bottomrule
\end{tabular}

\begin{tablenotes}[flushleft]
\footnotesize
\item[1] Based on the Circularity Gap Report 2025 \citep{circleeconomy2025}.
\item[2] Derived from the Outer Space Treaty \citep{unitednations1967}.
\end{tablenotes}

\end{threeparttable}
\end{table}

\newpage
\subsection{The three operational environments of the circular space economy}
\label{subsec:cse_conceptualization_three_operational_zones}

\begin{figure}[!htbp]
	\centering 
	\includegraphics[width=1.0\textwidth]{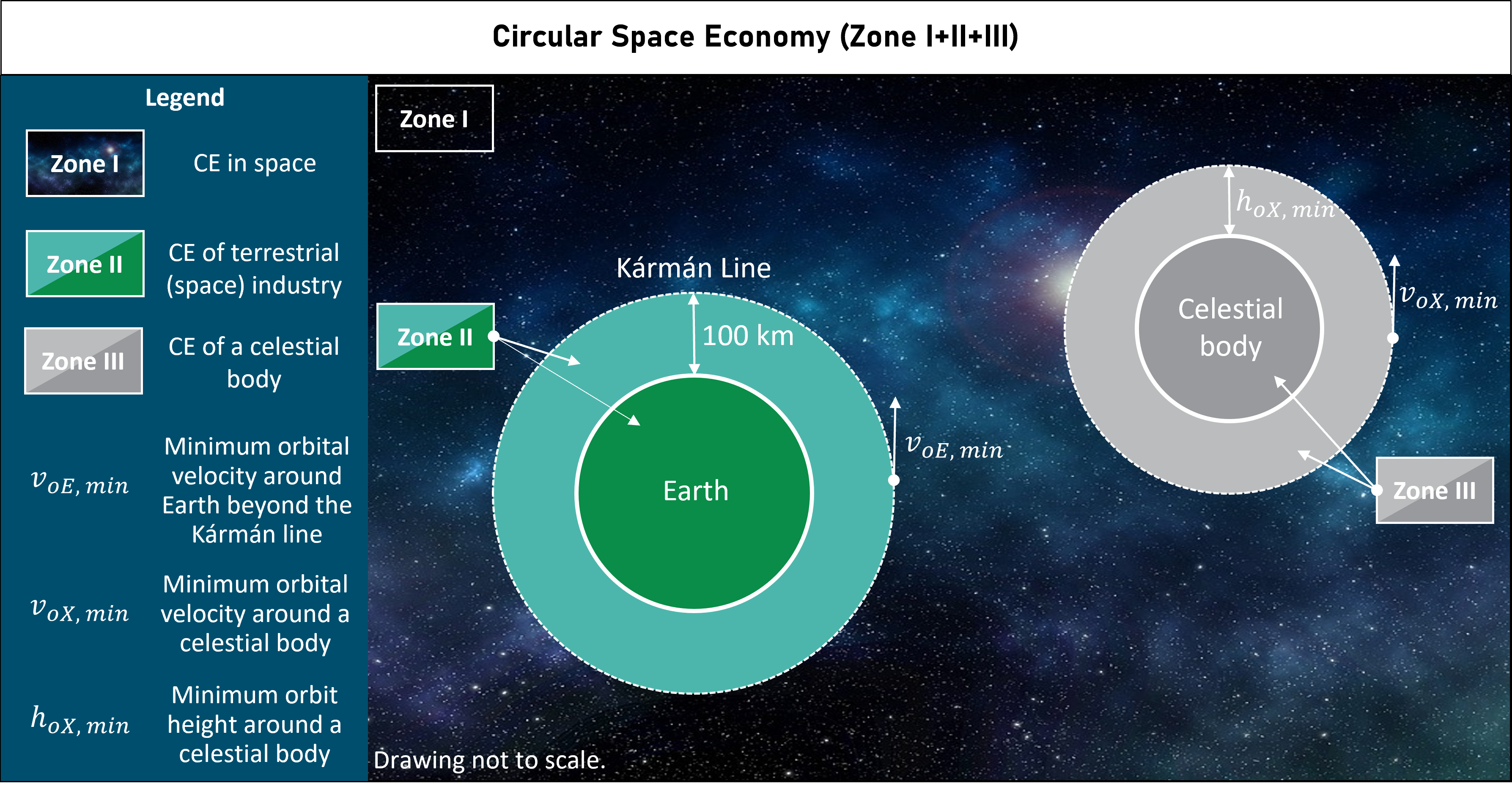}	
	\caption{The three operational environments (zones) of the CSE and associated terminology. The background image is sourced from Microsoft 365 stock images; all other elements were created by the authors.} 
	\label{fig:cse_overview}
\end{figure}

The Earth–space comparison motivated the distinction of three operational environments (zones) to establish a common scope and terminology for the CSE. This distinction is relevant for accounting and regulatory purposes, as circular activities may occur in different environments across the life cycle of a space system. It also aligns with ISO 59020:2024, where regional assessment levels are used to define system boundaries and can extend to planetary dimensions, including the atmosphere and Earth’s orbit \citep{iso2024b}. As shown in \ref{fig:cse_overview}, the CSE is therefore structured into three zones: (I) CE in space (CEiS), (II) CE of the terrestrial (space) industry, and (III) CE of celestial bodies beyond Earth.

Zone I covers circular activities involving objects that are physically in space. For Earth, this generally refers to objects beyond the Kármán Line, roughly 100 km above Earth’s surface \citep{dailey2024a}. To exclude objects that only temporarily cross this boundary, such as launch vehicle stages intended to return to Earth, the object must also reach at least minimum orbital velocity required for continuous orbital motion, while still accounting for elliptical orbits with perigees slightly below 100 km. In addition, this classification simplifies accounting, as atmospheric airspace is not part of Zone I by definition and related activities do not need to be tracked, except for launch vehicles capable of crossing the Kármán Line and reaching sustained orbital motion. This also excludes suborbital space tourism from Zone I.

Zone II captures circular activities of the terrestrial (space) industry, including design, manufacturing, testing, launch-related infrastructure, and ground-based recovery or recycling. Zone III extends the scope to celestial bodies beyond Earth, such as the Moon, Mars, or asteroids, where circular activities may involve local resources, infrastructure, and surface-based operations. Similar boundary requirements apply to activities around these bodies: objects that reach sustained orbital motion are accounted for in Zone I, while different celestial bodies themselves can be treated as sub-zones within Zone III. Resource exchanges between zones may include products, components, materials, labor, data, software, digital and non-digital services, energy, financial flows, and other value flows relevant to circularity; these include biological (renewable), technical (finite), and digital resources and may need to be tracked and managed \citep{geissdoerfer2017, ellenmcarthurfoundation2019, geissdoerfer2025}, depending on the purpose.

\subsection{The 10R Space Framework}
\label{subsec:cse_conceptualization_10r_space_framwork}

The CSE can be operationalized through circular actions that narrow, slow, and close resource loops. In this context, narrowing refers to reducing resource inputs in the pre-use phase, slowing to extending the lifetime and utility of products or components during use, and closing to recovering resources for further use in the post-use phase \citep{bocken2016a}. Where applicable, terminology follows ISO 59004:2024 and ISO 59020:2024, for instance through the term circular actions, which includes actions such as reduce, reuse, repair/maintain, refurbish, repurpose, remanufacture, recycle, valorization, and energy recovery \citep{iso2024b}; some of these are also described as resource management actions \citep{iso2024c}. In other literature, these actions are referred to as value retention options \citep{reike2018}, or simply R’s \citep{kirchherr2017, kirchherr2023}. Building on established frameworks \citep{potting2017, kirchherr2017, reike2018}, the 10R Space Framework adapts these circular actions to the specific conditions of space systems while maintaining a balance between simplicity and analytical depth, as shown in \ref{tab:10r_space_framework}.

Specified in Section \ref{subsec:cse_conceptualization_three_operational_zones}, these actions apply in principle to physical and non-physical, biological and non-biological resources. Their applicability and relative importance vary with the system in focus, its boundaries, and operational environment. For example, distributed systems in low Earth orbit (LEO) may provide different opportunities for circular actions than large monolithic space systems in geosynchronous Orbit (GEO). Generally, lower R-values indicate higher levels of circularity and should therefore be prioritized where technically, economically, and operationally feasible.

In particular, R1 Rethink should be highlighted, as it is often overlooked in current CSE discussions and, as indicated by an ongoing study, is not always considered part of the CSE. It promotes the more intensive use of space systems, for example through resource sharing, which is already applied where economically viable through ridesharing missions and payload hosting. A clear distinction is also required between R4 Repair, R5 Refurbish, and R6 Remanufacture, which may appear similar but differ substantially in scope, process intensity, and expected outcome, as specified in \ref{tab:10r_space_framework}. Finally, R10 Re-mine introduces a circular action that is particularly relevant in a space context, as it translates into the extraction of products, components, or materials from orbital sinks, such as debris clouds or graveyard orbits, followed by reintegration into lower-R circular actions. This makes re-mining systematically distinct from R8 Recycling, which focuses on planned material recovery and processing rather than the prior extraction of abandoned resources \citep{reike2018}.

The 10R Space Framework provides structured terminology and an implementation approach for circularity in space system design, operations, and life-cycle planning. In this work, the framework constitutes the final element of the CSE conceptualization needed to develop the structured CSE definition in the following section.

\begin{table}[ht]
\centering
\footnotesize
\caption{The 10R Space Framework, developed with inspiration from \citet{bocken2016a}, \citet{potting2017}, \citet{kirchherr2017}, \citet{reike2018}, and ISO 59004:2024, \textit{Circular economy -- Vocabulary, principles and guidance for implementation} \citep{iso2024c}. Terminology primarily follows the ISO standard where applicable. Circular actions with lower R-values indicate higher levels of circularity and should therefore be prioritized.}
\label{tab:10r_space_framework}
\begin{tblr}{
  width = 15cm, %\linewidth
  colspec = {Q[30]Q[123]Q[270]Q[270]},
  row{1} = {font=\bfseries},
  cell{2}{1} = {r=2}{c,t},
  cell{4}{1} = {r=5}{c,t},
  cell{9}{1} = {r=3}{c,t},
  hline{1,12} = {-}{0.08em},
  hline{2} = {1}{},
  hline{2} = {2-4}{0.03em},
  hline{4,9} = {-}{},
}
Loop & Circular action & Definition & Application\\
\begin{sideways}\parbox{1.72cm}{\centering Narrow \\ (pre-use phase)}\end{sideways} & R1 ~Rethink & Reconsider product or component use with the goal of using it more intensively. & Share infrastructure, share data, software, and computing power if possible, increase system uptime, improve operational efficiency.\\
 & R2 ~Reduce & Achieve the same functionality with fewer resources. & Improve design and manufacturing efficiency, minimize resource input, particularly virgin materials.\\
\begin{sideways}\parbox{2cm}{\centering Slow \\ (use phase)}\end{sideways} & R3 ~Reuse & Use products or components again for their original purpose without significant modification. & Reuse products or components through resale, redeployment, or storage for future use, supported by modular architectures, standardized interfaces, and forward and backward compatibility across spacecraft generations.\\
 & R4 ~Repair & Maintain a product or component so that it can be used in its original function. & Restore functionality through inspection, diagnostics, disassembly, replacement of faulty parts, and in-space or ground-based repair operations.\\
 & R5 ~Refurbish & Restore a product or component to similar, original, or improved quality and performance characteristics. & Cleaning, testing, part replacement, and software or hardware updates to improve performance.\\
 & R6 Remanufacture & Restore a product or component, through an industrial process, to a like-new condition and improved performance. & Restore products or components through inspection, standardized disassembly, diagnostics, replacement, reassembly, and qualification processes to achieve like-new condition and improved performance.\\
 & R7 ~Repurpose & Adapt a product or component for use in a different function than it was originally intended with or without making modifications. & Transform EoL products or components for alternative mission functions, such as inspection training objects, technology demonstration platforms, or shielding structures from spent upper stages and discarded structural elements.\\
\begin{sideways}\parbox{2cm}{\centering Close \\ (post-use phase)}\end{sideways} & R8 ~Recycle & Recover and process material to obtain the same or lower quality through activities such as collection, transport, sorting, cleaning and re-processing. & Use EoL products, components or materials, or production waste as feedstock for manufacturing new products and components in space and on ground.\\
 & R9 ~Recover energy & Generate useful energy from recovered resources. & Waste heat recovery and energy recovery from flammable materials not suitable for other circular actions (incineration).\\
 & R10 Re-mine & Mining or extraction of products, components, or materials from orbital sinks, followed by reintegration into lower-R circular actions. & Extract products, components, and materials from debris clouds, graveyard orbits, and discarded spacecraft units.\\
\end{tblr}
\end{table}

\section{Defining the circular space economy}
\label{sec:cse_definition}

\begin{figure}[ht!]
	\centering 
	\includegraphics[width=1.0\textwidth]{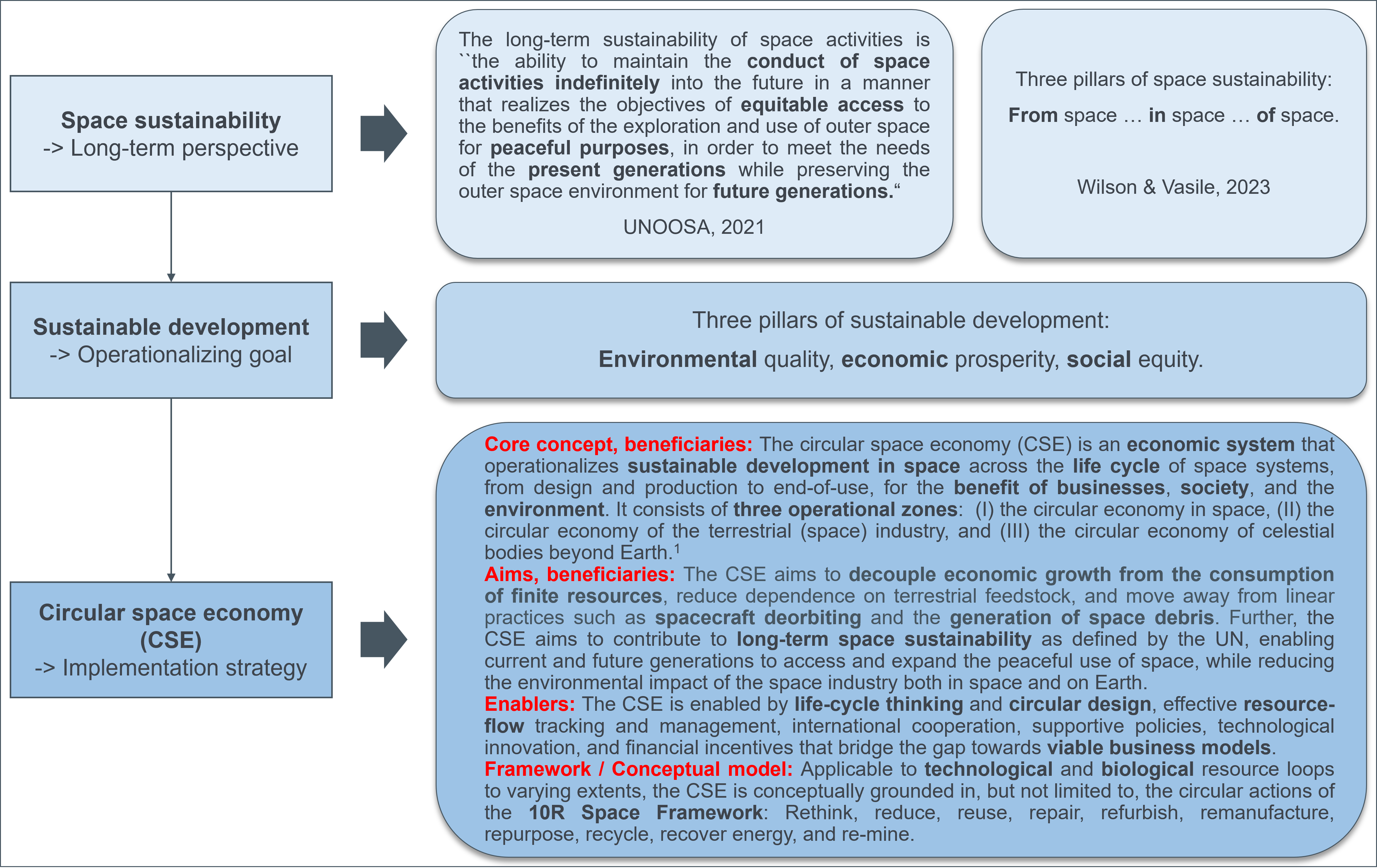}	
	\caption{Relationship between space sustainability, sustainable development, and the CSE, including a structured CSE definition.} 
	\label{fig:definition_overview}
\end{figure}

This section synthesizes the former analyses into a structured definition of the CSE. The investigation of CE definitions and CSE definition proposals determined the elements required for a definition, while the conceptualization section addressed missing aspects by examining Earth and space in a C(S)E context, distinguishing operational environments, clarifying terminology, and introducing the 10R Space Framework. As summarized in \ref{fig:definition_overview}, the definition positions the CSE as an implementation strategy for sustainable development in space and on Earth, understood through the pillars of environmental quality, economic prosperity, and social equity. Space sustainability provides the long-term perspective, based on the definitions by \citet{unitednationsofficeforouterspaceaffairsunoosa2021} and \citet{wilson2023}. It provides a basis for shared understanding across stakeholders and for further discussion. The resulting definition is structured along the coding dimensions developed in Section \ref{sec:methodology}, as follows.

\noindent\textbf{Core concept, beneficiaries:} The circular space economy (CSE) is an economic system that operationalizes sustainable development in space across the life cycle of space systems, from design and production to end-of-use, for the benefit of businesses, society, and the environment. It consists of three operational zones: (I) the circular economy in space, (II) the circular economy of the terrestrial (space) industry, and (III) the circular economy of celestial bodies beyond Earth.\footnote{An object is considered part of the circular economy in space once it reaches orbital velocity around a celestial body; for Earth, this implies sustained presence beyond the Kármán Line, which marks the regulatory boundary to space.}

\noindent\textbf{Aims, beneficiaries:} The CSE aims to decouple economic growth from the consumption of finite resources, reduce dependence on terrestrial feedstock, and move away from linear practices such as spacecraft deorbiting and the generation of space debris. Further, the CSE aims to contribute to long-term space sustainability as defined by the UN, enabling current and future generations to access and expand the peaceful use of space, while reducing the environmental impact of the space industry both in space and on Earth.

\noindent\textbf{Enablers:} The CSE is enabled by life-cycle thinking and circular design, effective resource-flow tracking and management, international cooperation, supportive policies, technological innovation, and financial incentives that bridge the gap towards viable business models.

\noindent\textbf{Framework / Conceptual model:} Applicable to technological and biological resource loops to varying extents, the CSE is conceptually grounded in, but not limited to, the circular actions of the 10R Space Framework: Rethink, reduce, reuse, repair, refurbish, remanufacture, repurpose, recycle, recover energy, and re-mine.

\section{Discussion and conclusion}
\label{sec:discussion_and_conclusion}

Since the beginning of the space age in 1957 \citep{nationalaeronauticsandspaceadministration2022}, space activities have largely followed a linear model in which objects are launched, operated, and eventually abandoned, disposed of in the Earth's atmosphere, or transferred to graveyard orbits. This model has contributed to resource losses, debris accumulation, environmental pollution, and operational risks. At the same time, reusable launch systems, large constellations, and new actors placing increasing pressure on the orbital environment. This situation will significantly intensify through the further expected decrease in launch costs once reusable heavy-lift launchers become fully operational. It is therefore questionable how long such environmentally unsustainable and economically inefficient practices can continue.

By conceptualizing the CSE, this paper provides the groundwork for the first structured CSE definition in the scientific literature. Hence, it establishes a basis for moving beyond the status quo towards a more circular model that supports sustainable growth, long-term cost reduction, and higher resource yields.

In detail, the work combines two analytical perspectives that have not yet been integrated in the scientific discourse: established CE definitions and CSE definition proposals. Further, broad and systematic gaps are identified and subsequently addressed, including the absence of a systematic distinction between in-space and terrestrial space-related activities, the lack of a life-cycle perspective, no differentiation between Earth and space as operational environments, and inconsistent terminology across sources. 

A central contribution of this work is that it broadens the current CSE discourse beyond individual circular actions. Existing CSE narratives are strongly centered on reuse, recycling, and in-orbit servicing, although the CE provides a wider range of actions for narrowing, slowing, and closing resource loops. Recycling remains an important part of circularity, but it is not the only, nor necessarily the preferred circular action. The 10R Space Framework therefore helps to position circular actions hierarchically and shows that circular actions with higher levels of circularity, such as rethink, reduce, reuse, repair, refurbish, and remanufacture, should be prioritized depending on the system, operational environment, and life-cycle phase.

However, several limitations need to be acknowledged. First, the coding approach required simplification. The decision to avoid interpretation of textual elements was made to improve the reliability of the results and reduce human error and personal influence in the coding process. Consequently, the strict focus on definitions may narrow the perspective on some studies, as discussed in Section \ref{sec:analysis}. In addition, some textual elements could plausibly be assigned to more than one dimension; to maintain clarity and comparability, elements were therefore classified according to their primary function. Yet, no case was identified in which a classification choice determined whether a dimension was triggered. Furthermore, the CSE literature is still emerging and distributed across peer-reviewed literature and grey literature, such as institutional documents and web-based sources, which do not always follow scientific standards. Although the selection aimed to capture the current definition landscape as comprehensively as possible, completeness cannot be guaranteed, primarily due to inconsistent terminology.

The proposed definition should therefore be understood as a foundation for further development, rather than a closure of the debate. Future work needs to further refine the CSE as a concept, examine stakeholder needs, priorities, uncertainties, and risks through direct engagement with relevant actors, and operationalize the CSE through case studies, particularly for large space systems and constellations. Addressing these gaps requires an interdisciplinary perspective, as the CSE depends not only on technical development but also on clear system boundaries, resource-flow tracking and management, stakeholder awareness, CE knowledge transfer, and effective implementation pathways. In the spirit of the Apollo 13 capsule communicator facing a complex problem: “One at a time, people. One at a time.”

\section*{Acknowledgments}

This work builds on prior research by the authors \citep{bahlmann2024, bahlmann2025}. The present project is/was supported by the National Research Fund, Luxembourg (grant number 17990314).

\appendix

\begin{landscape}
\section{Detailed analysis of established circular economy definitions}
\label{appendix_a_detailed_analysis_ce_definitions}

\scriptsize
\begin{longtable}{>{\raggedright\arraybackslash}p{1.5cm} >{\raggedright\arraybackslash}p{3.96cm} >{\raggedright\arraybackslash}p{2.925cm} >{\raggedright\arraybackslash}p{2.925cm} >{\raggedright\arraybackslash}p{2.6cm} >{\raggedright\arraybackslash}p{2.5cm} >{\raggedright\arraybackslash}p{2.5cm}}
\caption{Analysis of ten CE definitions, sorted by year of publication (if available) and selected according to the principles in Section \ref{sec:methodology}. Coding dimensions are partly adapted from \citet{kirchherr2023}.}
\label{tab:CE_defintions_analysis} \\
\toprule
\textbf{Source} & \textbf{Definition} & \textbf{Core concept} & \textbf{Aims} & \textbf{Enablers} & \textbf{Beneficiaries} & \textbf{Framework / Conceptual model} \\
\midrule
\endfirsthead

\multicolumn{6}{c}%
{{ Table \thetable\ (continued)}} \\
\toprule
\textbf{Source} & \textbf{Definition} & \textbf{Core concept} & \textbf{Aims} & \textbf{Enablers} & \textbf{Beneficiaries} & \textbf{Framework / Conceptual model} \\
\midrule
\endhead

\bottomrule
\endfoot

\cite{ellenmcarthurfoundation2013}; retrieved from \citet{kirchherr2017} &
A circular economy is restorative and regenerative by design, and aims to keep products, components, and materials at their highest utility and value at all times. The concept distinguishes between technical and biological cycles. As envisioned by the originators, a circular economy is a continuous positive development cycle that preserves and enhances natural capital, optimises resource yields, and minimises system risks by managing finite stocks and renewable flows. It works effectively at every scale. & - Restorative and regenerative by design \newline - Distinction between technical and biological cycles & - Keep products, components, and materials at their highest utility and value \newline - Continuous positive development cycle \newline - Preserves and enhances natural capital \newline - Minimize system risks \newline - Optimize resource yields \newline - Works effectively at every scale & - Design (restorative and regenerative) \newline - Management of finite stocks and renewable flows & - Not specifically mentioned & - Technical and biological cycles \newline - Not specifically mentioned, but it is associated to the Butterfly Diagram \\

\cite{geissdoerfer2017} &
... we define the Circular Economy as a regenerative system in which resource input and waste, emission, and energy leakage are minimised by slowing, closing, and narrowing material and energy loops. This can be achieved through long-lasting design, maintenance, repair, reuse, remanufacturing, refurbishing, recycling. & - Regenerative system \newline - Minimizing resource input and waste, emission, and energy leakage & - Slowing, closing, and narrowing material and energy loops & - Long-lasting design & - Not specifically mentioned & - 5R+ framework (long-lasting design, maintenance, repair, reuse, remanufacturing, refurbishing, recycling) \newline - Slowing, closing, and narrowing material and energy loops \\

\cite{murray2017} &
The Circular Economy is an economic model wherein planning, resourcing, procurement, production and reprocessing are designed and managed, as both process and output, to maximize ecosystem functioning and human well-being. & - Economic model \newline - Planning, resourcing, procurement, production and reprocessing are designed and managed & - Maximise ecosystem functioning \newline - Maximise human well-being & - Design and management of processes and outputs & - Humans & - Economic model (planning, resourcing, procurement, production and reprocessing are designed and managed) \\

\cite{kirchherr2017} &
A circular economy describes an economic system that is based on business models which replace the ‘end-of-life’ concept with reducing, alternatively reusing, recycling and recovering materials in production/distribution and consumption processes, thus operating at the micro level (products, companies, consumers), meso level (eco-industrial parks) and macro level (city, region, nation and beyond), with the aim to accomplish sustainable development, which implies creating environmental quality, economic prosperity and social equity, to the benefit of current and future generations. & - Economic system \newline - Based on business models \newline - Reducing, alternatively reusing, recycling and recovering materials in production/distribution and consumption processes & - Replace the ‘end-of-life’ concept \newline - Accomplish sustainable development \newline - Create environmental quality, economic prosperity and social equity  & - Business models \newline - Implementation at micro level (products, companies, consumers), meso level (eco-industrial parks) and macro level (city, region, nation and beyond) & - Current and future generations & - 4R framework (reducing, reusing, recycling, recovering) \newline - Micro, meso, and macro levels \\

\cite{unenvironmentprogramme2019} &
Acknowledging that a more circular economy, one of the current sustainable economic models, in which products and materials are designed in such a way that they can be reused, remanufactured[,] recycled or recovered and thus maintained in the economy for as long as possible, along with the resources of which they are made, and the generation of waste, especially hazardous waste, is avoided or minimized, and greenhouse gas emissions are prevented or reduced, can contribute significantly to sustainable consumption and production[.] & - Sustainable economic model \newline - Products and materials are designed to be reused, remanufactured, recycled, or recovered \newline - Avoid or minimise (especially hazardous) waste and greenhouse gas emissions & - Maintain products, materials, resources in the economy for as long as possible \newline - Contribute significantly to sustainable consumption and production & - Product and material design & - Not specifically mentioned & - Sustainable economic model \newline - 4R framework (reusing, remanufacturing, recycling, recovering) \\

\cite{kirchherr2023} &
The circular economy (CE) is a regenerative economic system which necessitates a paradigm shift to replace the 'end of life' concept with reducing, alternatively reusing, recycling, and recovering materials throughout the supply chain, with the aim to promote value maintenance and sustainable development, creating environmental quality, economic development, and social equity, to the benefit of current and future generations. It is enabled by an alliance of stakeholders (industry, consumers, policymakers, academia) and their technological innovations and capabilities. & - Regenerative economic system \newline - Reducing, alternatively reusing, recycling, and recovering materials throughout the supply chain & - Replace the 'end-of-life' concept \newline - Promote value maintenance \newline - Promote sustainable development \newline - Create environmental quality, economic development, and social equity & - Stakeholders (industry, consumers, policymakers, academia) \newline - Technological innovations and capabilities & - Current and future generations & - 4R framework (reducing, reusing, recycling, recovering) \\

\cite{iso2024c} &
Economic system that uses a systemic approach to maintain a circular flow of resources, by recovering, retaining or adding to their value, while contributing to sustainable development. & - Economic system \newline
- Systemic approach \newline - Maintain a circular flow of resources & - Contribute to sustainable development & - Recovering, retaining, or adding value to resources & - Not specifically mentioned & - Not specifically mentioned \\

\cite{mckinseyandcompany2024} &
Circularity presents an alternative to the linear model. In a circular economy, resources can be used over and over again, often for the same or similar purposes. Three major principles govern a circular economy: 1. Preserve and enhance natural capital (the world’s stock of natural assets) by controlling finite resources and balancing the flow of renewable resources. 2. Optimize resource yields by circulating products, components, and materials in use at the highest possible levels at all times. 3. Make the system more effective by eliminating unintended negative consequences, like air and water pollution. & - Alternative to linear model \newline - System perspective \newline - Preserve and enhance natural capital \newline - Optimize resource yields \newline - Make the system more effective & - Use resources over and over again (often for the same or similar purposes) \newline - Preserve and enhance natural capital (the world’s stock of natural assets) \newline - Optimize resource yields \newline - Make the system more effective & - Control finite resources and balance the flow of renewable resources \newline - Circulate products, components, and materials in use at the highest possible levels at all times \newline - Eliminate unintended negative consequences (air and water pollution) & - Not specifically mentioned & - Three major principles govern a circular economy: \newline - Preserve and enhance natural capital \newline - Optimize resource yields \newline - Make the system more effective \\

\cite{ellenmcarthurfoundation2025} &
The circular economy is a system where materials never become waste and nature is regenerated. In a circular economy, products and materials are kept in circulation through processes like maintenance, reuse, refurbishment, remanufacture, recycling, and composting. The circular economy tackles climate change and other global challenges, like biodiversity loss, waste, and pollution, by decoupling economic activity from the consumption of finite resources. The circular economy is based on three principles, driven by design: Eliminate waste and pollution; Circulate products and materials (at their highest value); Regenerate nature. Underpinned by a transition to renewable energy and materials, the circular economy is a resilient system that is good for business, people, and the environment. & - Resilient system, driven by design \newline - Eliminate waste and pollution \newline  - Circulate products and materials (at their highest value) \newline - Regenerate nature & - Prevent materials from becoming waste and regenerate nature \newline - Tackle climate change, biodiversity loss, waste, and pollution & - Decouple economic activity from the consumption of finite resources \newline - Design \newline - Renewable energy and materials & - Business \newline - People \newline - Environment & - 4R+ framework (maintenance, reuse, refurbishment, remanufacture, recycling, composting)  \newline - Not specifically mentioned, but it is associated to the Butterfly Diagram \\

\cite{europeanunionn.d.} &
A circular economy is a system which maintains the value of products, materials and resources in the economy for as long as possible, and minimises the generation of waste. This means a system where products are reused, repaired, remanufactured or recycled. & - System perspective \newline - Value retention of products, materials and resources within economy \newline - Minimise waste generation & - Lifetime extension of resources within the economy \newline - Reduce waste generation & - Not specifically mentioned & - Not specifically mentioned & - 4R framework (reuse, repair, remanufacture, recycle) \\
\end{longtable}
\end{landscape}

\begin{landscape}
\section{Detailed analysis of circular space economy definition proposals}
\label{appendix_b_detailed_analysis_cse_definition_proposals}

\scriptsize
\begin{longtable}{>{\raggedright\arraybackslash}p{1.5cm} >{\raggedright\arraybackslash}p{4.7cm} >{\raggedright\arraybackslash}p{2.837cm} >{\raggedright\arraybackslash}p{2.837cm} >{\raggedright\arraybackslash}p{2.837cm} >{\raggedright\arraybackslash}p{1.8cm} >{\raggedright\arraybackslash}p{2.4cm}}
\caption{Analysis of 13 CSE definition proposals, sorted by year of publication (if available) and selected according to the principles in Section \ref{sec:methodology}. Coding dimensions are partly adapted from \citet{kirchherr2023}.}
\label{tab:CSE_defintions_analysis} \\
\toprule
\textbf{Source} & \textbf{Definition proposal} & \textbf{Core concept} & \textbf{Aim} & \textbf{Enabler} & \textbf{Beneficiary} & \textbf{Framework / Conceptual model} \\
\midrule
\endfirsthead

\multicolumn{6}{c}%
{{ Table \thetable\ (continued)}} \\
\toprule
\textbf{Source} & \textbf{Definition proposal} & \textbf{Core concept} & \textbf{Aim} & \textbf{Enabler} & \textbf{Beneficiary} & \textbf{Framework / Conceptual model} \\
\midrule
\endhead

\bottomrule
\endfoot

\cite{wilson2023} & 
This vision of a circular space economy signifies the use and re-use of space resources, including in orbit servicing, repair, manufacturing and recycling. It also ensures the responsible management of space debris, resources, environmental impact mitigation and equitable access to space-based opportunities. In this sense, it should firstly prioritise the minimisation and mitigation of space debris, aiming for zero proliferation through stringent debris mitigation measures, active debris removal initiatives and responsible end-of-life practices. Secondly, the framework should strive to eliminate all unnecessary adverse impacts on the Earth environment by minimising emissions and other negative environmental consequences associated with space activities. This entails adopting eco-design principles, promoting transparency/openness in reporting and implementing sustainable operational practices. Lastly, the framework should ensure zero unsustainable use of resources by fostering a circular economy approach, promoting in-orbit servicing, repair, manufacturing and recycling by means of efficient resource utilisation. Through the integration of modular design principles, the disassembly and repurposing of satellites and space structures become achievable objectives ... Such a scenario could transform previously undesirable or problematic materials into valuable feedstock for future manufacturing processes, paving the way for the achievement of the net-zero and circular space economy vision via the development of new and effective space sustainability business models. & - Use and re-use of space resources \newline - In-orbit servicing, repair, manufacturing, and recycling \newline - Responsible management of space debris and resources \newline - Environmental impact mitigation \newline - Equitable access to space-based opportunities & - Equitable access to space-based opportunities \newline - Zero proliferation of space debris \newline - Eliminate unnecessary adverse impacts on Earth's environment \newline - Transform previously undesirable or problematic materials into valuable feedstock for future manufacturing processes \newline - Achievement of the net-zero and circular space economy vision & - Stringent debris mitigation measures \newline - Active debris removal initiatives \newline - Responsible end-of-life
practices \newline - Minimising emissions and other negative environmental consequences associated with space activities \newline - Eco-design principles \newline - Transparency / openness in reporting and implementing
sustainable operational practices \newline - Efficient resource utilisation \newline - Modular design principles (enables the disassembly and repurposing
of satellites and space structures) \newline - New and effective business models & - Not specifically mentioned & - Proposes the creation of an international framework, suggesting requirements: \newline - Minimisation and mitigation of space debris \newline - Elimination of all unnecessary adverse impacts on the Earth environment \newline - Zero unsustainable use of resources \\

\cite{leonard2023} & 
One potential knock-on effect of the rapid implementation of in-orbit servicing solutions would be the shifting of the space economy from linear to circular. A circular space economy would aim to maintain maximum utility of its products and materials \citep{brennan2020}. & - Opposed to linear space economy \newline - Maintain maximum utility of products and materials & - Maintain maximum utility of products and materials & - In-orbit servicing & - Not specifically mentioned & - Not specifically mentioned \\

\cite{europeanspaceagency2023} &
... ESA is encouraging the implementation of a ‘circular economy’ in space that ensures long-term orbital sustainability through an ecosystem of in-orbit servicing, in-orbit assembly, in-orbit manufacturing, and eventually in-orbit recycling. ... The implementation of a Space Circular Economy could play an important role in guaranteeing the sustainability of the orbits, maximising the usage of space assets (reduction of costs) and protecting the Earth’s environment by limiting the exploitation of raw materials on-ground and lowering the number of satellites launches and re-entries. & - Long-term orbital sustainability \newline - In-orbit servicing \newline - In-orbit assembly \newline - In-orbit manufacturing \newline - In-orbit recycling & - Ensure long-term orbital sustainability \newline - Maximising the usage of space assets (reduction of costs) \newline - Protecting the Earth’s environment (limiting the exploitation of raw materials on-ground; lowering the number of satellites launches and re-entries) & - TBD & - Not specifically mentioned & - Not specifically mentioned \\

\cite{jah2024a} &
The circular space economy draws inspiration from the broader concept of a circular economy, which aims to minimize waste and maximize resource efficiency. At its core, this approach advocates for the development and operation of reusable and recyclable satellites, spacecraft, and space infrastructure. By integrating circular design principles, such as modular and interoperable components, into satellite engineering, the generation of space debris can be significantly reduced. & - Reusable and recyclable satellites, spacecraft, and space infrastructure & - Minimize waste \newline - Maximize resource efficiency \newline - Reduce generation of space debris & - Modular and interoperable components by design & - Not specifically mentioned & - Not specifically mentioned \\

\cite{bahlmann2024a} &
Focusing on the space segment, the CSE replaces the traditional end-of-life concept with an end-of-use approach. It aims to avoid waste creation during all mission phases, providing a new life by design for all involved technical and biological resources after their intended use phase. CSE serves as a performance enhancement strategy to achieve long-term space sustainability. & - Replace end-of-life concept by end-of-use approach \newline - New life by design for technical and biological resources after use phase & - Avoid waste creation during all mission phases \newline - Performance enhancement strategy to achieve long-term space sustainability & - Design & - Not specifically mentioned & - Not specifically mentioned \\

\cite{turner2024} &
A space economy in which novel methods of design and managing space systems allows systems, subsystems, components, and materials to remain in orbit and be refurbished or re-used using an ecosystem of advanced in-orbit servicing techniques. This approach will preserve the space environment for future generations while meeting current economic needs. & - Systems, subsystems, components, and materials remain in orbit \newline - Refurbishment and re-use in orbit & - Preserve the space environment \newline - Meet current economic needs & - Novel methods of design and managing space systems \newline - Advanced in-orbit servicing techniques & - Future generations & - Not specifically mentioned \\

\cite{dailey2024} &
The circular space economy represents a transformative shift in addressing the pressing challenge of space debris while fostering sustainable economic growth in both space and on Earth. & - Transformative shift in addressing space debris & - Foster sustainable economic growth in space and on Earth & - Not specifically mentioned & - Not specifically mentioned & - Not specifically mentioned \\

\cite{noaaofficeofspacecommerce2025} &
The circular space economy approach aims to utilize space-based resources sustainably by minimizing waste and maximizing the reuse and recycling of materials in space operations. As humanity expands its presence beyond Earth, this model becomes vital for reducing the need for costly resupply missions and mitigating the environmental impact of space activities. & - Minimize waste \newline - Maximize reuse and recycling of materials in space operations & - Utilize space-based resources sustainably \newline - Reduce need for costly resupply missions \newline - Mitigate environmental impact of space activities & - Not specifically mentioned & - Not specifically mentioned & - Not specifically mentioned \\

\cite{bennett2025} & 
To encourage conscientious use of finite terrestrial and orbital resources, governments should promote a circular space economy, which embraces waste reduction, reuse, and recycling, through policy and technology innovations. & - Conscientious use of finite terrestrial and orbital resources \newline - Waste reduction \newline - Reuse \newline - Recycling & - Conscientious use of finite terrestrial and orbital resources & - Governments \newline - Policy  \newline - Technology innovations & - Not specifically mentioned & - Not specifically mentioned \\

\cite{d-orbit2026} & 
Space circular economy means creating a sustainable economy in space by reducing waste, reusing resources, and recycling materials. This involves designing space missions, spacecraft, and space habitats with the goal of achieving a closed-loop system where waste products are repurposed or recycled to create new products. This can help address the challenges of sustainable resource use in space; as human activity in space expands, the amount of waste generated and resources consumed will also increase, which could lead to environmental degradation and the depletion of finite resources. In the future satellites will be built and manufactured in space, raw material will come from asteroids mining and from recycling of existing assets in space; when we will reach that point, we will be very close to a space circular economy. By adopting circular economy principles in space, we can create a sustainable system that can support long-term human habitation and exploration, which is an integral part of our Vision. & - Sustainable economy in space \newline - Sustainable system \newline - Closed-loop system where waste products are repurposed or recycled to create new products & - Reduce waste \newline - Reuse resources \newline - Recycle materials \newline - Address the challenges of sustainable resource use in space \newline - Address the increasing amount of waste generated and resources consumed will increase, since human activity in space will expand \newline - Avoid environmental degradation \newline - Avoid depletion of finite resources \newline - Support long-term human habitation and exploration & - Design of space missions, spacecraft, and space habitats \newline - In-space satellite building and manufacturing \newline - Raw materials from asteroid mining recycling of existing assets in space \newline - Adopting circular economy principles in space & - Humans & - Not specifically mentioned \\

\cite{saadetturnbull2026} & 
A circular space economy is one in which space systems are designed and operated in a manner that enables those systems and their subsystems, components, and materials to remain in use through an ecosystem of space-based servicing and manufacturing techniques. This approach seeks to minimize waste, optimize resource use, and preserve the space environment for future generations, while supporting the long-term sustainability of space activities in all its aspects. & - Space systems are designed and operated to remain in use \newline - System, subsystem, component, and material levels & - Minimize waste \newline - Optimize resource use \newline - Preserve the space environment \newline - Support the long-term sustainability of space activities in all its aspects & - Space-based servicing \newline - Space-based manufacturing techniques & - Future generations & - Not specifically mentioned \\

\cite{prismsustainabilitydirectoryn.d.} &
The Circular Space Economy signifies a strategic reorientation of resource management and value creation within the domain of space activities, aligning with global imperatives for planetary stewardship and sustainable development. It represents a systemic shift towards responsible utilization of space assets and resources, addressing both terrestrial environmental concerns and the long-term viability of human activities beyond Earth. This conceptual framework supports broader discussions on resource governance, waste reduction, and the ethical stewardship of the space environment, positioning space exploration and utilization within a planetary carrying capacity context. & - Strategic reorientation of resource management and value creation \newline - Systemic shift towards responsible utilization of space assets and resources \newline - Conceptual framework, positioning space exploration and utilization within a planetary carrying capacity context & - Planetary stewardship \newline - Sustainable development \newline - Address terrestrial environmental concerns \newline - Long-term viability of human activities beyond Earth & - Not specifically mentioned & - Not specifically mentioned & - Not specifically mentioned (``conceptual framework" used as synonym for CSE) \\

\cite{wealthformulan.d.} &
Space circular economy is using, repairing and recycling in orbit hardware, instead of launching new satellite or letting others become debris. This could reduce the demand for raw materials, reduce satellite launches, and preserve Earth’s environment. & - Using, repairing and recycling in-orbit hardware instead of launching new satellites or letting others become debris & - Reduce demand for raw materials \newline - Reduce satellite launches \newline - Preserve Earth’s environment & - Not specifically mentioned & - Not specifically mentioned & - Not specifically mentioned \\
\end{longtable}
\end{landscape}

%% If you have bibdatabase file and want bibtex to generate the
%% bibitems, please use
%%
\bibliographystyle{elsarticle-harv} 
\bibliography{bibliography}

@misc{nationalaeronauticsandspaceadministration2022,
	title = {Dawn of the {Space} {Age} - {NASA}},
	url = {https://www.nasa.gov/history/dawn-of-the-space-age/},
	language = {en-US},
	urldate = {2026-05-11},
	author = {{National Aeronautics and Space Administration}},
	year = {2022},
	note = {{Accessed} on 11 May 2026},
}

@techreport{saadetturnbull2026,
	title = {Exploring the {Circular} {Economy} in {Space}},
	copyright = {Creative Commons Attribution 4.0 International},
	url = {https://figshare.com/articles/online_resource/_b_Exploring_the_Circular_Economy_in_Space_b_/31285906},
	doi = {10.6084/M9.FIGSHARE.31285906},
	urldate = {2026-04-27},
	institution = {figshare},
	author = {Saadet Turnbull, Dr. Beril and Turner, Calum},
	month = feb,
	year = {2026},
	note = {Artwork Size: 6509445 Bytes
Pages: 6509445 Bytes},
}

@article{winans2017,
	title = {The history and current applications of the circular economy concept},
	volume = {68},
	issn = {1364-0321},
	url = {https://www.sciencedirect.com/science/article/pii/S1364032116306323},
	doi = {10.1016/j.rser.2016.09.123},
	urldate = {2026-04-27},
	journal = {Renewable and Sustainable Energy Reviews},
	author = {Winans, K. and Kendall, A. and Deng, H.},
	month = feb,
	year = {2017},
	pages = {825--833},
}

@article{kessler1978,
	title = {Collision frequency of artificial satellites: {The} creation of a debris belt},
	volume = {83},
	copyright = {This paper is not subject to U.S. copyright. Published in 1978 by the American Geophysical Union.},
	issn = {2156-2202},
	shorttitle = {Collision frequency of artificial satellites},
	url = {https://onlinelibrary.wiley.com/doi/abs/10.1029/JA083iA06p02637},
	doi = {10.1029/JA083iA06p02637},
	language = {en},
	number = {A6},
	urldate = {2026-04-25},
	journal = {Journal of Geophysical Research: Space Physics},
	author = {Kessler, Donald J. and Cour-Palais, Burton G.},
	year = {1978},
	note = {\_eprint: https://agupubs.onlinelibrary.wiley.com/doi/pdf/10.1029/JA083iA06p02637},
	pages = {2637--2646},
}

@techreport{circleeconomy2026,
	address = {Amsterdam},
	title = {The circularity gap report 2026: {The} value gap},
	url = {https://dashboard.circularity-gap.world/report/2026/cgr-2026-overview},
	language = {en},
	urldate = {2026-04-16},
	institution = {Circle Economy},
	author = {{Circle Economy}},
	year = {2026},
}

@article{bocken2016a,
	title = {Product design and business model strategies for a circular economy},
	volume = {33},
	issn = {2168-1015, 2168-1023},
	url = {http://www.tandfonline.com/doi/full/10.1080/21681015.2016.1172124},
	doi = {10.1080/21681015.2016.1172124},
	language = {en},
	number = {5},
	urldate = {2026-04-14},
	journal = {Journal of Industrial and Production Engineering},
	author = {Bocken, Nancy M. P. and De Pauw, Ingrid and Bakker, Conny and Van Der Grinten, Bram},
	month = jul,
	year = {2016},
	pages = {308--320},
}

@misc{unitednations1967,
	type = {Treaty},
	title = {Treaty on {Principles} {Governing} the {Activities} of {States} in the {Exploration} and {Use} of {Outer} {Space}, including the {Moon} and {Other} {Celestial} {Bodies}},
	url = {https://www.unoosa.org/oosa/en/ourwork/spacelaw/treaties/introouterspacetreaty.html},
	urldate = {2026-04-10},
	publisher = {United Nations Office for Outer Space Affairs (UNOOSA)},
	author = {{United Nations}},
	year = {1967},
}

@techreport{circleeconomy2025,
	address = {Amsterdam},
	title = {The circularity gap report 2025},
	url = {https://pdf.circularity-gap.world/?report=CGR_Global_2025_Report_0c90048033&page=1},
	language = {en},
	urldate = {2026-04-10},
	institution = {Circle Economy},
	author = {{Circle Economy}},
	year = {2025},
}

@article{2026,
	title = {Space capacity-based metric to rank in orbit collision risk},
	volume = {77},
	issn = {0273-1177},
	url = {https://www.sciencedirect.com/science/article/pii/S0273117726001274},
	doi = {10.1016/j.asr.2026.01.077},
	language = {en-US},
	number = {6},
	urldate = {2026-03-17},
	journal = {Advances in Space Research},
	publisher = {Pergamon},
	month = mar,
	year = {2026},
	pages = {7054--7066},
}

@article{bonifazi2026,
	title = {Circular space economy: {A} review of innovative sustainable waste management},
	issn = {0734-242X, 1096-3669},
	shorttitle = {Circular space economy},
	url = {https://journals.sagepub.com/doi/10.1177/0734242X251415529},
	doi = {10.1177/0734242X251415529},
	language = {en},
	urldate = {2026-03-10},
	journal = {Waste Management \& Research: The Journal for a Sustainable Circular Economy},
	author = {Bonifazi, Giuseppe and D’Adamo, Idiano and Grosso, Chiara and Palmieri, Roberta and Zorpas, Antonis A.},
	month = feb,
	year = {2026},
	pages = {0734242X251415529},
}

@misc{d-orbit2026,
	title = {Sustainability {\textbar} {D}-{Orbit}},
	url = {https://www.dorbit.space/sustainability},
	urldate = {2026-03-05},
	author = {{D-Orbit}},
	year = {2026},
}

@article{yang2025,
	title = {Resource and material efficiency in the circular space economy},
	volume = {0},
	issn = {3051-2948},
	url = {https://www.cell.com/chem-circularity/abstract/S3051-2948(25)00001-5},
	doi = {10.1016/j.checir.2025.100001},
	language = {English},
	number = {0},
	urldate = {2026-02-09},
	journal = {Chem Circularity},
	publisher = {Elsevier},
	author = {Yang, Zhilin and Liu, Lirong and Xing, Lei and Amara, Adam and Xuan, Jin},
	month = dec,
	year = {2025},
}

@article{schultz2026,
	title = {The {Circular} {Space} {Economy}: {Review}, {Conceptualization}, and {Research} {Agenda} for {Narrowing}, {Slowing}, and {Closing} the {Loop} in {Orbit}},
	volume = {4},
	shorttitle = {The {Circular} {Space} {Economy}},
	doi = {10.55845/joce-2026-41216},
	journal = {Journal of Circular Economy},
	author = {Schultz, Felix Carl},
	month = feb,
	year = {2026},
}

@misc{iso2024b,
	edition = {1},
	title = {Circular economy — {Measuring} and assessing circularity performance ({ISO} 59020:2024)},
	shorttitle = {{ISO} 59020:2024},
	url = {https://www.iso.org/standard/80650.html},
	language = {en},
	urldate = {2026-02-03},
	author = {{ISO}},
	year = {2024},
}

@misc{iso2024c,
	edition = {1},
	title = {Circular economy — {Vocabulary}, principles and guidance for implementation ({ISO} 59004:2024)},
	shorttitle = {{ISO} 59004:2024},
	url = {https://www.iso.org/standard/80648.html},
	language = {en},
	urldate = {2026-02-03},
	author = {{ISO}},
	year = {2024},
}

@article{geissdoerfer2025,
	title = {Conceptualizing {Circular} {Ecosystems}: {An} {Analysis} of 45 {Definitions}},
	issn = {0964-4733, 1099-0836},
	shorttitle = {Conceptualizing {Circular} {Ecosystems}},
	url = {https://onlinelibrary.wiley.com/doi/10.1002/bse.70242},
	doi = {10.1002/bse.70242},
	language = {en},
	urldate = {2025-10-30},
	journal = {Business Strategy and the Environment},
	author = {Geissdoerfer, Martin and Kanda, Wisdom and Kirchherr, Julian},
	month = oct,
	year = {2025},
	pages = {bse.70242},
}

@inproceedings{bahlmann2025,
	address = {Esch-sur-Alzette, Luxembourg},
    title = {Towards {Defining} the {Circular} {Space} {Economy}},
	booktitle = {Doctoral {Programme} in {Computer} {Science} and {Computer} {Engineering} PhD {Day}},
	url = {https://www.researchgate.net/doi/10.13140/RG.2.2.29784.56320},
	doi = {10.13140/RG.2.2.29784.56320},
	language = {en},
	urldate = {2025-10-07},
	author = {Bahlmann, Jonas and Saidani, Michael and Stoll, Enrico and Hein, Andreas M.},
	year = {2025},
    pages = {1--1},
}

@misc{europeanspaceagency2025,
	title = {{ESA} {Space} {Environment} {Report} 2025},
	url = {https://www.esa.int/Space_Safety/Space_Debris/ESA_Space_Environment_Report_2025},
	language = {en},
	urldate = {2025-10-01},
	author = {{ESA}},
	year = {2025},
}

@inproceedings{bahlmann2024a,
	address = {Milan, Italy},
	title = {[{Presentation}] {Space} and the {Circular} {Economy}: {Exploring} {Expert} {Perceptions}},
	shorttitle = {[{Presentation}] {Space} and the {Circular} {Economy}},
	doi = {10.13140/RG.2.2.28200.23045},
	language = {English},
	urldate = {2025-09-26},
	booktitle = {Proceedings of the 75th {International} {Astronautical} {Congress}},
	publisher = {International Astronautical Federation},
	author = {Bahlmann, Jonas and Saidani, Michael and Franzese, Vittorio and Stoll, Enrico and Hein, Andreas},
	month = oct,
	year = {2024},
    pages = {1--22},
	note = {IAC-24,D1,1,1,x90997},
}

@inproceedings{dailey2024a,
	address = {Milan, Italy},
	title = {Integrating {Orbital} {Carrying} {Capacity} into {International} {Policy} {Constructs}: {Leveraging} {Best} {Practices} from {Aviation}'s {Risk}-{Based} {Norms}.},
	isbn = {979-8-3313-1211-4},
	shorttitle = {Integrating {Orbital} {Carrying} {Capacity} into {International} {Policy} {Constructs}},
	url = {http://www.proceedings.com/078360-0175.html},
	doi = {10.52202/078360-0175},
	urldate = {2025-09-22},
	booktitle = {22nd {IAA} {Symposium} on {Space} {Debris}},
	publisher = {International Astronautical Federation (IAF)},
	author = {Dailey, Nathaniel and Stilwell, Ruth and Malekos Smith, Zhanna and McKnight, Darren},
	year = {2024},
	pages = {1806--1818},
}

@article{bennett2025,
	title = {Orbital debris requires prevention and mitigation across the satellite life cycle},
	volume = {4},
	copyright = {2025 The Author(s)},
	issn = {2731-3395},
	url = {https://www.nature.com/articles/s44172-025-00430-5},
	doi = {10.1038/s44172-025-00430-5},
	language = {en},
	number = {1},
	urldate = {2025-07-22},
	journal = {Communications Engineering},
	publisher = {Nature Publishing Group},
	author = {Bennett, Mia M.},
	month = may,
	year = {2025},
	pages = {95},
}

@article{brennan2020,
	title = {The {Orbital} {Circular} {Economy} {Framework}: {Emblematic} {Evidence} from the {Space} {Industry}},
	volume = {8},
	issn = {2186-6961},
	url = {https://ualresearchonline.arts.ac.uk/id/eprint/18579/1/KMRVol.8-6.Brennan%20and%20Vecchi%27s%20paper_0417.pdf#:~:text=future,a%20completely%20new%20level%20in},
	language = {en},
	journal = {Kindai Management Review},
	author = {Brennan, Louis and Vecchi, Alessandra},
	year = {2020},
	pages = {81--93},
}

@inproceedings{dailey2024,
	address = {Milan, Italy},
	title = {Leveraging a {Circular} {Economy} for {Space} {Sustainability}: {Government} {Roles} and {Economic} {Impacts}},
	shorttitle = {Leveraging a {Circular} {Economy} for {Space} {Sustainability}},
	url = {http://www.proceedings.com/078380-0020.html},
	doi = {10.52202/078380-0020},
	language = {en},
	urldate = {2025-07-21},
	booktitle = {37th {IAA} {Symposium} on {Space} {Policy}, {Regulations} and {Economics}},
	publisher = {International Astronautical Federation (IAF)},
	author = {Dailey, Nathaniel and Groesbeck, Thomas and Toner, Kevin and Engel, Rozlyn and Malekos Smith, Zhanna and Steinke, Lee},
	year = {2024},
	pages = {170--180},
}

@misc{wealthformulan.d.,
	title = {Funding the {Future}: {Innovative} {Solutions} for {Space} {Debris} {Removal}},
	shorttitle = {Funding the {Future}},
	url = {https://www.wealthformula.com/blog/funding-the-future-innovative-solutions-for-space-debris-removal/},
	language = {en-US},
	urldate = {2025-07-21},
	journal = {Wealth Formula},
	author = {{Wealth Formula}},
}

@misc{prismsustainabilitydirectoryn.d.,
	title = {Circular {Space} {Economy}},
	url = {https://prism.sustainability-directory.com/area/circular-space-economy/},
	language = {en-GB},
	urldate = {2025-07-21},
	journal = {Prism → Sustainability Directory},
	author = {{Prism Sustainability Directory}},
}

@misc{noaaofficeofspacecommerce2025,
	title = {Circular {Space} {Economy} {Seminar} on {February} 13 – {Office} of {Space} {Commerce}},
	url = {https://space.commerce.gov/circular-space-economy-seminar-on-february-13/},
	language = {en-US},
	urldate = {2025-07-21},
	journal = {Circular Space Economy Seminar on February 13},
	author = {{NOAA Office of Space Commerce}},
	month = feb,
	year = {2025},
}

@misc{jah2024a,
	title = {Space {Environmentalism} - {Toward} a {Circular} {Economy} {Approach} for {Orbital} {Space} {\textbar} {National} {Academies}},
	url = {https://www.nationalacademies.org/news/2024/04/space-environmentalism-toward-a-circular-economy-approach-for-orbital-space},
	urldate = {2025-07-21},
	journal = {NAE Perspectives},
	author = {Jah, Moriba},
	month = apr,
	year = {2024},
}

@inproceedings{turner2024,
	address = {Milan, Italy},
	title = {Enabling a {Space} {Circular} {Economy} by 2050},
	url = {http://www.proceedings.com/078372-0006.html},
	doi = {10.52202/078372-0006},
	urldate = {2025-07-21},
	booktitle = {{IAF} {Space} {Systems} {Symposium}},
	publisher = {International Astronautical Federation (IAF)},
	author = {Turner, Calum and Caiazzo, Antonio and Tormena, Enrico and Soares, Tiago},
	month = oct,
	year = {2024},
	pages = {63--67},
}

@article{murray2017,
	title = {The {Circular} {Economy}: {An} {Interdisciplinary} {Exploration} of the {Concept} and {Application} in a {Global} {Context}},
	volume = {140},
	copyright = {http://www.springer.com/tdm},
	issn = {0167-4544, 1573-0697},
	shorttitle = {The {Circular} {Economy}},
	url = {http://link.springer.com/10.1007/s10551-015-2693-2},
	doi = {10.1007/s10551-015-2693-2},
	language = {en},
	number = {3},
	urldate = {2025-07-17},
	journal = {Journal of Business Ethics},
	publisher = {Springer Science and Business Media LLC},
	author = {Murray, Alan and Skene, Keith and Haynes, Kathryn},
	month = feb,
	year = {2017},
	pages = {369--380},
}

@misc{unenvironmentprogramme2019,
	title = {Resolution adopted by the {United} {Nations} {Environment} {Assembly} on 15 {March} 2019},
	url = {https://documents.un.org/doc/undoc/gen/k19/010/42/pdf/k1901042.pdf},
	urldate = {2025-07-17},
	publisher = {United Nations},
	author = {{UN Environment Programme}},
	year = {2019},
}

@misc{mckinseyandcompany2024,
	title = {What is circularity?},
	url = {https://www.mckinsey.com/featured-insights/mckinsey-explainers/what-is-circularity},
	urldate = {2025-07-17},
	author = {{McKinsey \& Company}},
	year = {2024},
}

@misc{europeanunionn.d.,
	title = {Circular economy - {EUR}-{Lex}},
	url = {https://eur-lex.europa.eu/legal-content/EN/TXT/?uri=legissum:circular_economy},
	language = {en},
	urldate = {2025-07-08},
	author = {{European Union}},
	note = {Doc ID: circular\_economy
Doc Sector: other
Doc Title: Circular economy
Doc Type: other
Usr\_lan: en},
}

@misc{ellenmcarthurfoundation2019,
	title = {Circular economy systems diagram},
	url = {https://www.ellenmacarthurfoundation.org},
	author = {{Ellen McArthur Foundation}},
	month = feb,
	year = {2019},
}

@misc{ellenmcarthurfoundation2025,
	title = {Circular economy introduction},
	url = {https://www.ellenmacarthurfoundation.org/topics/circular-economy-introduction/overview},
	language = {en-GB},
	urldate = {2025-04-14},
	author = {{Ellen McArthur Foundation}},
	year = {2025},
    note = {{Accessed} on 14 April 2025},
}

@misc{ellenmcarthurfoundation2026,
	title = {Circular economy introduction},
	url = {https://www.ellenmacarthurfoundation.org/topics/circular-economy-introduction/overview},
	language = {en-GB},
	urldate = {2026-03-13},
	author = {{Ellen McArthur Foundation}},
	year = {2026},
    note = {{Accessed} on 13 March 2026},
}

@misc{ellenmcarthurfoundation2013,
	title = {Circular {Economy} {Overview}},
	url = {https://www.ellenmacarthurfoundation.org/circular-economy/overview/concept},
	author = {{Ellen McArthur Foundation}},
	year = {2013},
}

@inproceedings{bahlmann2024,
	address = {Milan, Italy},
	title = {Space and the {Circular} {Economy}: {Exploring} {Expert} {Perceptions}},
    url = {https://www.researchgate.net/publication/385710397_Space_and_the_Circular_Economy_Exploring_Expert_Perceptions},
	doi = {10.52202/078372-0001},
    pages = {1--16},
	language = {English},
	booktitle = {Proceedings of the 75th {International} {Astronautical} {Congress}},
	publisher = {International Astronautical Federation},
	author = {Bahlmann, Jonas and Saidani, Michael and Franzese, Vittorio and Stoll, Enrico and Hein, Andreas},
	month = oct,
	year = {2024},
	note = {IAC-24,D1,1,1,x90997},
}

@misc{ellenmcarthurfoundationn.d.,
	title = {The circular economy in detail},
	url = {https://www.ellenmacarthurfoundation.org/the-circular-economy-in-detail-deep-dive},
	language = {en-GB},
	urldate = {2024-10-18},
	author = {{Ellen McArthur Foundation}},
}

@incollection{manickam2019,
	title = {{3Rs} and circular economy},
	copyright = {https://www.elsevier.com/tdm/userlicense/1.0/},
	isbn = {978-0-08-102630-4},
	url = {https://linkinghub.elsevier.com/retrieve/pii/B9780081026304000042},
	doi = {10.1016/B978-0-08-102630-4.00004-2},
	language = {en},
	urldate = {2024-10-18},
	booktitle = {Circular {Economy} in {Textiles} and {Apparel}},
	publisher = {Elsevier},
	author = {Manickam, Parthiban and Duraisamy, Gopalakrishnan},
	year = {2019},
	pages = {77--93},
}

@inproceedings{wohlin2014,
	address = {London England United Kingdom},
	title = {Guidelines for snowballing in systematic literature studies and a replication in software engineering},
	isbn = {978-1-4503-2476-2},
	url = {https://dl.acm.org/doi/10.1145/2601248.2601268},
	doi = {10.1145/2601248.2601268},
	language = {en},
	urldate = {2024-09-27},
	booktitle = {Proceedings of the 18th {International} {Conference} on {Evaluation} and {Assessment} in {Software} {Engineering}},
	publisher = {ACM},
	author = {Wohlin, Claes},
	month = may,
	year = {2014},
	pages = {1--10},
}

@article{reike2018,
	title = {The circular economy: {New} or {Refurbished} as {CE} 3.0? — {Exploring} {Controversies} in the {Conceptualization} of the {Circular} {Economy} through a {Focus} on {History} and {Resource} {Value} {Retention} {Options}},
	volume = {135},
	issn = {09213449},
	shorttitle = {The circular economy},
	url = {https://linkinghub.elsevier.com/retrieve/pii/S0921344917302756},
	doi = {10.1016/j.resconrec.2017.08.027},
	language = {en},
	urldate = {2024-09-24},
	journal = {Resources, Conservation and Recycling},
	author = {Reike, Denise and Vermeulen, Walter J.V. and Witjes, Sjors},
	month = aug,
	year = {2018},
	pages = {246--264},
}

@techreport{unitednationsofficeforouterspaceaffairsunoosa2021,
	address = {Vienna, Austria},
	title = {Guidelines for the {Long}-term {Sustainability} of {Outer} {Space} {Activities} of the {Committee} on the {Peaceful} {Uses} of {Outer} {Space}},
	url = {https://www.unoosa.org/documents/pdf/PromotingSpaceSustainability/Publication_Final_English_June2021.pdf},
	language = {en},
	institution = {United Nations (UN)},
	author = {{UNOOSA}},
	year = {2021},
}

@book{potting2017,
	title = {Circular {Economy}: {Measuring} innovation in the product chain},
	shorttitle = {Circular {Economy}},
	url = {http://www.pbl.nl/sites/default/files/cms/publicaties/pbl-2016-circular-economy-measuring-innovation-in-product-chains-2544.pdf},
	author = {Potting, José and Hekkert, M.P. and Worrell, Ernst and Hanemaaijer, Aldert},
	month = jan,
	year = {2017},
}

@article{kirchherr2023,
	title = {Conceptualizing the {Circular} {Economy} ({Revisited}): {An} {Analysis} of 221 {Definitions}},
	volume = {194},
	issn = {0921-3449},
	shorttitle = {Conceptualizing the {Circular} {Economy} ({Revisited})},
	url = {https://www.sciencedirect.com/science/article/pii/S0921344923001374},
	doi = {10.1016/j.resconrec.2023.107001},
	urldate = {2024-08-27},
	journal = {Resources, Conservation and Recycling},
	author = {Kirchherr, Julian and Yang, Nan-Hua Nadja and Schulze-Spuentrup, Frederik and Heerink, Maarten J. and Hartley, Kris},
	month = jul,
	year = {2023},
	pages = {107001},
}

@article{geissdoerfer2017,
	title = {The {Circular} {Economy} – {A} new sustainability paradigm?},
	volume = {143},
	issn = {0959-6526},
	url = {https://www.sciencedirect.com/science/article/pii/S0959652616321023},
	doi = {10.1016/j.jclepro.2016.12.048},
	urldate = {2024-08-13},
	journal = {Journal of Cleaner Production},
	author = {Geissdoerfer, Martin and Savaget, Paulo and Bocken, Nancy M. P. and Hultink, Erik Jan},
	month = feb,
	year = {2017},
	pages = {757--768},
}

@article{kirchherr2017,
	title = {Conceptualizing the circular economy: {An} analysis of 114 definitions},
	volume = {127},
	issn = {0921-3449},
	shorttitle = {Conceptualizing the circular economy},
	url = {https://www.sciencedirect.com/science/article/pii/S0921344917302835},
	doi = {10.1016/j.resconrec.2017.09.005},
	urldate = {2024-02-16},
	journal = {Resources, Conservation and Recycling},
	author = {Kirchherr, Julian and Reike, Denise and Hekkert, Marko},
	month = dec,
	year = {2017},
	pages = {221--232},
}

@misc{europeanspaceagency2023,
	title = {Enabling a {Space} {Circular} {Economy} by 2050},
	url = {https://indico.esa.int/event/450/contributions/8930/attachments/5891/9829/Space%20Circular%20Economy%20White%20Paper%20Issue%201.pdf},
	urldate = {2024-01-31},
	publisher = {European Space Agency (ESA)},
	author = {{ESA}},
	year = {2023},
}

@article{leonard2023,
	title = {Viability of a circular economy for space debris},
	volume = {155},
	issn = {0956-053X},
	url = {https://www.sciencedirect.com/science/article/pii/S0956053X22005104},
	doi = {10.1016/j.wasman.2022.10.024},
	urldate = {2024-01-31},
	journal = {Waste Management},
	author = {Leonard, Ryan and Williams, Ian D.},
	month = jan,
	year = {2023},
	pages = {19--28},
}

@article{wilson2023,
	title = {The space sustainability paradox},
	volume = {423},
	issn = {09596526},
	url = {https://linkinghub.elsevier.com/retrieve/pii/S0959652623030275},
	doi = {10.1016/j.jclepro.2023.138869},
	language = {en},
	urldate = {2024-01-30},
	journal = {Journal of Cleaner Production},
	author = {Wilson, Andrew Ross and Vasile, Massimiliano},
	month = oct,
	year = {2023},
	pages = {138869},
}

%% else use the following coding to input the bibitems directly in the
%% TeX file.

%%\begin{thebibliography}{00}

%% \bibitem[Author(year)]{label}
%% For example:

%% \bibitem[Aladro et al.(2015)]{Aladro15} Aladro, R., Martín, S., Riquelme, D., et al. 2015, \aas, 579, A101

%%\end{thebibliography}

\end{document}